\documentstyle[preprint,aps]{revtex}

\begin{document}
\draft
\title{Quasiparticle current in ballistic $NcS^{\prime}S$ junctions}
\author{A.A.Golubov$^{1,2}$ and M.Yu.Kupriyanov$^{1,3}$}
\address{$^1$Institute of Thin Film and Ion Technology,
Research Centre J\"ulich (KFA)
\\ D-52425 J\"ulich, Germany
\\ $^2$Permanent address: Institute of Solid State Physics,
142432 Chernogolovka, Russia
\\ $^3$Permanent address:
Nuclear Physics Institute, Moscow State University, 119899 GSP Moscow, Russia}
\maketitle

\begin{abstract}
Nonstationary properties of ballistic constrictions $NcS^{\prime}S$ with
disordered $S^{\prime}S$ electrodes are analyzed theoretically. Amplitudes
of Andreev and normal reflections at the constriction are related to the
solutions of a stationary Green function problem in an inhomogeneous $%
S^{\prime}S$ electrode in a dirty limit. This provides a generalization of
the model of Blonder, Tinkham and Klapwijk for a spatially inhomogeneous
case. The relation between quasiparticle current in $NcS^{\prime}S$
junctions and energy spectrum of a $S^{\prime}S$ proximity sandwich is found
for arbitrary parameters of $S^{\prime}$ and $S$ materials and of $%
S^{\prime}S$ interface.\\
\end{abstract}

\newpage

\section{Introduction}

Tunnel junctions with high critical current density are presently a subject
of extensive experimental investigation (see \cite{Klein} and references
therein). For such junctions the transparency of a tunnel barrier is not
small, and therefore they do not fulfill the conditions of a standard tunnel
theory. As was shown in \cite{Klein}, barriers in $Nb-AlO_x-Nb$ junctions
with high $J_c$ are likely to be a series of constrictions, each having
rather large transparency. Therefore, a small ballistic $ScS$ constriction
is a suitable starting point to discuss more complicated models for high J$%
_c $junctions.

The properties of $NcS$ constrictions are well understood in the framework
of the model of Blonder, Tinkham and Klapwijk (BTK) \cite{BTK}. In their
approach, current through a constriction is fully determined by the
amplitudes of normal and Andreev reflections at the $NS$ interface. The
model assumes that both N and S metals are in thermal equilibrium and that
both are spatially homogeneous. Later, the model was further developed by
Klapwijk et al, and Octavio et al \cite{OTBK,Flens} (KBT and OTBK models) to
treat $ScS$ constriction, in which the subharmonic structure on $I-V$ curves
was explained to be due to multiple Andreev reflections at the constriction.
Whereas the condition of thermal equilibrium is generally fulfilled for the
constriction geometry, the other condition of spatial homogeneity of the
superconducting electrodes is less general. An important case of an
inhomogeneous system is the $SS^{\prime }cS^{\prime }S$ junction, where $%
S^{\prime }$ is a superconductor with $T_c^{\prime }<T$, ($T_c^{\prime }=0$
corresponds tothe particular case of a normal metal).

Previous work on the generalization of the BTK approach to account for
spatial inhomogeneity was started by Van Son et al \cite{vKemp}. Andreev
reflection in the $NcN^{\prime }S$ system (where $N^{\prime }$ is a normal
metal) was considered for a gradual variation of a pair potential near the
normal metal - superconductor interface. To model Andreev reflection from
the $N^{\prime }$ region with proximity induced superconductivity, the
existence of a spatially dependent pair potential $\Delta _{n^{\prime }}(x)$
was assumed in $N^{\prime }$. However, whereas the Cooper pair density in a
normal region is indeed nonzero due to the proximity effect \cite{prox}, the
pair potential $\Delta _{n^{\prime }}(x)=0$, if $T_c^{\prime }=0$. Therefore
a more consistent approach is needed to modify the Andreev amplitudes in the
$NcN^{\prime }S$ sandwich in comparison to the BTK case of $NcS$. Such a
theory is necessary, in particularly, to interpret experiments on point
contact spectroscopy of proximity systems, like the one performed recently
on a bilayer consisting of doped $Si$ backed with superconducting $Nb$ \cite
{Hesl}.

It will be shown in this paper that the existence of a pair potential $%
\Delta _{n^{\prime }}(x)$ in $N^{\prime }$ is not a necessary condition for
Andreev reflection at the $NcN^{\prime }$ boundary. As is known, the
spectrum in a normal region is modified due to the proximity effect, namely,
bound states exist in $N^{\prime }$ layer at energies below the energy gap
of a superconductor, $\Delta _s$ \cite{bound-st,En-Wohl}. The energy of the
lowest bound state corresponds to an energy gap $\Delta _{gn^{\prime }}$
induced in $N^{\prime }$, which for a thin enough layer is close to $\Delta
_s$. Therefore it is clear qualitatively that in $NcN^{\prime }S$ systems
Andreev reflection processes take place not only at the $N^{\prime }S$
boundary, but also at the $NcN^{\prime }$ boundary, because at $E<\Delta
_{gn^{\prime }}$ a quasiparticle can not penetrate into $N^{\prime }$. This
result however is not directly evident from Bogolubov de Gennes (BdG)
equations, where the pair potential $\Delta _{n^{\prime }}(x)$ plays the
role of an effective scattering potential for nondiagonal scattering. To
find such a potential for $N^{\prime }S$ sandwiches, one should either solve
a complete three-layer problem in the BdG equations, taking into account
both $NN^{\prime }$ and $N^{\prime }S$ boundaries simultaneously, or to use
the microscopic Green functions approach. For a disordered $S^{\prime }S$
system the first approach would require knowledge of the full scattering
matrix, while the second one is much more straightforward, and we shall use
it in this paper to find the nondiagonal potential for tunneling into the $%
S^{\prime }S$ bilayer and to calculate the quasiparticle current for the $%
NcS^{\prime }S$ contact.

Previously, a microscopic approach based on Green functions formalism was
used to study the properties of $NcS$ and $NcN^{\prime }S$ microcontacts
without impurity scattering (clean limit). For the $NcS$ case, Zaitsev \cite
{Zai-84} has derived boundary conditions for the quasiclassical Eilenberger
equations at the contact interface and has calculated the current through
the contact at arbitrary transparency of the interface, thus providing a
microscopic derivation of BTK model. Independently, the same derivation was
done by Arnold \cite{Arn} with another method of retarded Green functions.
As a particular case, properties of clean $NIN^{\prime }S$ contacts with
arbitrary barrier transparency were considered in \cite{Arn}. More recently,
detailed calculations for the $NcN^{\prime }S$ case were done in the
framework of the Arnold model in \cite{chiara} with the additional
assumption of small transparency barrier between $N^{\prime }$ and $S$
layers.

In a case of strong disorder (dirty limit) nonequilibrium aspects got a lot
of attention recently. In \cite{Zai-90,Volk,VZK,Been} the conductivity of a
dirty $NN^{\prime }S$ contact was studied in the model, where $N$ and $S$
electrodes are reservuars with fixed electrical potentials whereas the
disordered contact region $N^{\prime }$ is a one-dimensional bridge in
nonequilibrium. Such a situation can be realized in disordered $NS$ contacts
when the $N^{\prime }$ layer simulates the interface region of the order of
the inelastic mean free path. An enhancement of the zero bias conductance $%
\sigma (0)$ was predicted in this system.

In the present paper we consider a physically different situation: a
ballistic constriction $NcS^{\prime }S$ of the size smaller than the mean
free path having disordered electrodes. In this case the potential drop
takes place at the constriction, and therefore the electrodes are in thermal
equilibrium. The ballistic condition is important. It should be compared
with the opposite limit of disordered constriction of a size larger than a
mean free path. The latter case was studied theoretically by Artemenko,
Volkov and Zaitsev \cite{AVZ} for $ScS$ constrictions and more recently by
Volkov \cite{Volk-94} for $NcN^{\prime }S$ constrictions with large
transparency. It was obtained, in particularly, that due to disorder no
conductance doubling is present at zero bias $V=0$. As is shown in this
paper, the conductance of the ballistic $NcN^{\prime }S$ constriction is
different in many respects as zero bias conductance doubling is being
present for large transparency. The influence of the finite transparency of
a constriction is also studied. Physically, our approach is closely related
to the BTK one. We will give expressions for coefficients of Andreev and
ordinary electron reflections at the ballistic constriction and their
relation to the energy spectrum of the disordered $SS^{\prime }$ system that
is investigated in terms of Green functions. As a result, the simplicity and
direct physical meaning of the BTK solutions are combined with the approach
based on the selfconsistent solution of a stationary dirty limit Green
functions problem.

\section{The Model}

Let us consider the boundary between $N$ and $S^{\prime }S$ as a small
constriction of size $a\ll \min (l_n,l_{s^{\prime }})$, where $%
l_n,l_{s^{\prime }}$ are mean free paths of $N$ and $S^{\prime }$. The
constriction is characterized by a transmission coefficient

\begin{equation}
D=\frac{4v_{1x}v_{2x}}{(v_{1x}+v_{2x})^2+4H^2}
\end{equation}
where $H$ is strength of repulsive potential $H\delta (x)$ located at the $%
NS^{\prime }$ interface and $v_{1x},v_{2x}$ are the components of the Fermi
velocities of $N$ and $S^{\prime }$ normal to the interface, respectively.
We assume that $S^{\prime }$ and $S$ metals are in the dirty limit $%
l_{s^{^{\prime }},s}\ll \xi _{s^{^{\prime }},s},$ whereas in $N$ there is no
limitation on the mean free path.

Let us follow the BTK notations. In the BdG equation formalism, the
excitations are represented by a vector $\psi =\left(
\begin{array}{c}
f(x) \\
g(x)
\end{array}
\right) $, where $f(x)$ describes electron-like excitations in a
superconductor, and $g(x)$ describes hole-like (time-reversed) excitations.
The BdG equations have the form:

\begin{equation}
\begin{array}{c}
i\hbar
\frac{\partial f}{\partial t}=\left( -\frac{\hbar ^2\nabla ^2}{2m}-\mu
(x)+V(x)\right) f(x)+\Delta (x)g(x,t) \\ i\hbar \frac{\partial g}{\partial t}%
=-\left( -\frac{\hbar ^2\nabla ^2}{2m}-\mu (x)+V(x)\right) g(x)+\Delta
(x)f(x,t)
\end{array}
\end{equation}
where $\mu (x),\Delta (x)$ and $V(x)$ are the electrochemical potential, the
pair potential, and the ordinary potential, respectively.

In the absence of impurity scattering, the transmission and reflection
coefficients for a quasiparticle incident from clean $N$ to clean $S$, are
found in the BTK model by considering incoming, reflected and transmitted
waves near the $NS$ boundary:

\begin{equation}
\psi _{inc}=\left(
\begin{array}{c}
1 \\
0
\end{array}
\right) e^{iq_1^{+}x},\text{ }\hbar q_{1,2}^{\pm }=\sqrt{2m_{1,2}(\mu \pm
\epsilon )}
\end{equation}

\begin{equation}
\begin{array}{c}
\psi _{refl}=a\left(
\begin{array}{c}
0 \\
1
\end{array}
\right) e^{iq_1^{-}x}+b\left(
\begin{array}{c}
1 \\
0
\end{array}
\right) e^{-iq_1^{+}x} \\
\psi _{trans}=c\left(
\begin{array}{c}
u_0 \\
v_0
\end{array}
\right) e^{iq_2^{+}x}+d\left(
\begin{array}{c}
v_0 \\
u_0
\end{array}
\right) e^{-iq_2^{-}x}
\end{array}
\end{equation}

\begin{equation}
1-v_0^2=u_0^2=\frac 12(1+\sqrt{\epsilon ^2-\Delta ^2}/\epsilon )
\end{equation}
where $\Delta $ is a the bulk energy gap of $S$, $m_{1,2\text{ }}$are
effective masses of the contacting metals and $\epsilon $ is the
quasiparticle energy.

Matching these solutions at the NS boundary, one can find the
energy-dependent Andreev reflection coefficient $A(\epsilon )$ and normal
reflection coefficient $B(\epsilon )$:

\begin{equation}
\begin{array}{c}
A(\epsilon )=aa^{*}=
\frac{\left| \eta \right| ^2}{(1+Z^2(1-\left| \eta \right| ^2))^2}, \\
B(\epsilon )=bb^{*}=\frac{Z^2(1+Z^2)(1-\left| \eta \right| ^2)^2}{%
(1+(1-\left| \eta \right| ^2)Z^2)^2}
\end{array}
\end{equation}

\begin{equation}
\eta (\epsilon )=\frac{v_0(\epsilon )}{u_0(\epsilon )}=\frac{\Delta /\sqrt{%
\epsilon ^2-\Delta ^2}}{1+E/\sqrt{\epsilon ^2-\Delta ^2}}
\end{equation}
where Z is related to normal transmission coefficient $D$ by: $%
(1+Z^2)=D^{-1}.$ It is important to note that both $A(\epsilon )$ and $%
B(\epsilon )$ are controlled by one energy-dependent parameter $\eta
(\epsilon )$, $\left| \eta \right| ^2$ being simply the ratio of
probabilities for the excitation to be in hole-like, $\left| v_0\right| ^2,$
or in electron-like, $\left| u_0\right| ^2,$ states.

The solutions (5) for $u_0,v_0$ are applicable for a clean homogeneous BCS
superconductor, and they are not valid in spatially inhomogeneous dirty $%
S^{\prime }S$ sandwiches. To find the coefficients $A(\epsilon )$ and $%
B(\epsilon )$ in the latter case one needs to calculate $\eta (\epsilon )$
as a solution of Gor'kov equations (GE) in the $SS^{\prime }$ system. For
the ballistically clean constriction we can take the advantage that at a
length scale smaller than the electron mean free path the solution of the GE
can be written in the form of a combination of plane waves.

The GE in the region near the constriction, $S^{\prime },$ have the form
\cite{AGD}:

\begin{equation}
\begin{array}{c}
\left\{ 2m_2(\epsilon ^{\prime }(x)+\mu )+
\frac{\partial ^2}{\partial x^2}\right\} G_\epsilon (x,x^{\prime
})+2m_2\Delta ^{\prime }(x)F_\epsilon ^{*}(x,x^{\prime })=\delta
(x-x^{\prime }) \\ \left\{ 2m_2(\epsilon ^{\prime }(x)-\mu )-\frac{\partial
^2}{\partial x^2}\right\} F_\epsilon (x,x^{\prime })+2m_2\Delta ^{\prime
}(x)G_\epsilon (x,x^{\prime })=0
\end{array}
\end{equation}
Here $E^{\prime }$ and $\Delta ^{\prime }(x)$ are the energy and pair
potential renormalized by impurity scattering according to
\begin{equation}
\epsilon ^{\prime }(x)=\epsilon +i\left\langle G_\epsilon (x,x)\right\rangle
/2\tau ,\text{ }\Delta ^{\prime }(x)=\Delta (x)+\left\langle F_\epsilon
(x,x)\right\rangle /2\tau ,
\end{equation}
$\tau =(2\pi cV^2N(0))^{-1}$ is scattering time, $c$ and $V$ are impurity
concentration and scattering potential, respectively. The brackets $%
\left\langle ...\right\rangle $ denote angle averaging, and $G_\epsilon
(x,x^{\prime })$ and $F_\epsilon (x,x^{\prime })$ are the normal and
anomalous Green functions in energy representation.

The pair potential in (8) and (9) is given by the selfconsistency equation:

\begin{equation}
\Delta (x)=gT\sum_{w_n}F_{\epsilon =-iw_n}(x,x)
\end{equation}
where $g$ is the coupling constant and $\omega _n=\pi T(2n+1)$ is the
Matsubara frequency. Note that in the particular case of $T_c^{\prime }=0$
the pair potential in $S^{\prime }\ $is zero, $\Delta (x)=0$, whereas $%
F_\epsilon (x,x)$ is finite.

We note that $\left\langle G_\epsilon (x,x)\right\rangle $ and $\left\langle
F_\epsilon (x,x)\right\rangle $ are quasiclassical Green functions of a
dirty superconductor. They obey diffusion-like equations \cite{Eliash,LO}
with the boundary conditions derived in \cite{KL}.

Let us consider the solutions of the GE in the $S^{\prime }$ region at
distances less than $\xi _{s^{\prime }}$ from the constriction. In this
region one can neglect variations of $\Delta ^{\prime }(x)$ and $\epsilon
^{\prime }(x)$ and to write the solutions of the linearized equations as a
combination of plane waves:

\begin{equation}
\left(
\begin{array}{c}
G_\epsilon (x,x^{\prime }) \\
F_\epsilon (x,x^{\prime })
\end{array}
\right) =C(x^{\prime })\left(
\begin{array}{c}
g(x) \\
f(x)
\end{array}
\right) e^{iq_2^{+}x}+D(x^{\prime })\left(
\begin{array}{c}
f(x) \\
g(x)
\end{array}
\right) e^{-iq_2^{-}x},
\end{equation}
where the slowly varying functions $g(x),$ $f(x)$ determine amplitudes of
electron like and hole like excitations. Substitution of the solution
eq.(11) to the GE leads to the following linear system of equations for $g(x)
$ and $f(x)$:
\begin{equation}
L\left(
\begin{array}{c}
g(x) \\
f(x)
\end{array}
\right) \equiv \left(
\begin{array}{cc}
2m_2\left[ \epsilon +i\left\langle G_\epsilon (x)\right\rangle /2\tau +\mu
\right] -q_2^2 & 2m_2\left[ \Delta +\left\langle F_\epsilon (x)\right\rangle
/2\tau \right]  \\
2m_2\left[ \Delta +\left\langle F_\epsilon (x)\right\rangle /2\tau \right]
& 2m_2\left[ \epsilon +i\left\langle G_\epsilon (x)\right\rangle /2\tau -\mu
\right] +q_2^2
\end{array}
\right) \left(
\begin{array}{c}
g(x) \\
f(x)
\end{array}
\right) =0.
\end{equation}
Here $q_2^{\pm }$ is determined by the dispersion relation $DetL=0$, which
in the dirty limit $\Delta \tau \ll 1$ leads to the following result

\begin{equation}
q_2^{\pm }=\sqrt{2m_2\left[ \mu \pm i\frac{\sqrt{\left\langle G_\epsilon
(x)\right\rangle ^2+\left\langle F_\epsilon (x)\right\rangle ^2}}{2\tau }%
\right] }.
\end{equation}

Eq.(11) describes transmitted electron and transmitted hole waves in the $%
S^{\prime }$ region near the constriction. Note that, in general, reflected
electron and reflected hole waves are also present in $S^{\prime }$ due to
impurity scattering. However, the condition of a ballistic constriction, $%
a\ll l_{s^{\prime }}$, made it possible to neglect these waves in eq.(11),
because under this condition waves in $S^{\prime }$ are scattered
diffusively away from the contact, and there is a small probability for a
wave, scattered at the distance $l_{s^{\prime }}$ from the constriction, to
reach it again.

To find transmission and reflection probabilities for a given problem, one
should compare the solutions (4) and (11) at distances smaller than mean
free path $l_{s^{\prime }}$ from the constriction. In this region one can
neglect the small energy terms in (3) and (13) in comparison with chemical
potential $\mu $, i.e. set $q_2^{+}=q_2^{-}=\sqrt{2m_2\mu }$ in the phases
of all transmitted waves. Then, to find the probabilities $A(\epsilon )$ and
$B(\epsilon )$ one should find the ratio $f(x)/g(x)$. For the dirty limit we
obtain from eq.(12):

\begin{equation}
\frac{f(x)}{g(x)}=\frac{i\left\langle F_\epsilon (x)\right\rangle }{%
1+\left\langle G_\epsilon (x)\right\rangle }.
\end{equation}
Finally, substituting $\eta (\epsilon )$ in eqs.(6),(7) by $f(x)/g(x)$ from
eq.(14) and using the normalization condition for the Green functions $%
\left\langle G_\epsilon ^2\right\rangle +\left\langle F_\epsilon
^2\right\rangle =1,$ we find that the Andreev and normal reflection
coefficients are directly related to the local energy spectrum of $S^{\prime
}$ near the constriction:

\begin{equation}
A(\epsilon )=\frac{\left| \left\langle F_\epsilon (0+)\right\rangle \right|
^2}{\left| 1+2Z^2+\left\langle G_\epsilon (0+)\right\rangle \right| ^2},
\end{equation}

\begin{equation}
B(\epsilon )=\frac{4Z^2(1+Z^2)}{\left| 1+2Z^2+\left\langle G_\epsilon
(0+)\right\rangle \right| ^2},
\end{equation}
where $\left\langle G_\epsilon (0+)\right\rangle $ and $\left\langle
F_\epsilon (0+)\right\rangle $ are quasiclassical Green functions in the
vicinity of the contact in $S^{\prime }$. The expressions (15) and (16) are
the central result of the paper. They generalize the corresponding BTK
relations \cite{BTK} for a spatially inhomogeneous case. For the latter case
the Green functions are given by $\left\langle G_\epsilon (0+)\right\rangle
=-i\epsilon /\sqrt{\Delta _0^2-\epsilon ^2},$ $\left\langle F_\epsilon
(0+)\right\rangle =\Delta _0/\sqrt{\Delta _0^2-\epsilon ^2}$ , and it is
easy to check, that the BTK relations follow from eqs.(15), (16).

The coefficients $A(\epsilon )$ and $B(\epsilon )$ given by eqs.(15,16)
fully determine the quasiparticle current through $NcS^{\prime }S$ contact.
Namely under the condition of thermal equilibrium in both electrodes the
current through the constriction is expressed via $A(\epsilon )$ and $%
B(\epsilon )$ in complete analogy with the BTK result for the $NcS$
constriction \cite{BTK} :
\begin{equation}
I(V)=\frac{R_0^{-1}}{\left\langle 1+Z^2\right\rangle }\int_{-\infty
}^{+\infty }\left[ f_0(\epsilon +eV)-f_0(\epsilon )\right] \left[
1+A(\epsilon )-B(\epsilon )\right] d\epsilon
\end{equation}
where $R_0=\left[ 2N_1(0)Se^2v_{F1}\right] ^{-1}$ is the Sharvin resistance,
$S$ is contact area, $N_1(0)$ and $v_{F1}$ are the density of states per
spin and Fermi velocity of $N$ electrode, respectively (we have taken $%
v_{F1}<v_{F2}$). $f_0$ is the Fermi distribution function and the brackets $%
\left\langle ...\right\rangle $ denote angular averaging. The relations (15)
and (16) show, that in the dirty limit local information is given by point
contact measurements. In particular, the local density of states near the
constriction is given in the usual way as $N(\epsilon )=Re\left\{
\left\langle G_\epsilon (0+)\right\rangle \right\} $. It is seen however
from eqs.(15)-(17), that generally (for arbitrary $Z$) the current is not
determined solely by $N(\epsilon )$ like in a tunnel theory, but rather a
crossover to the limit of a tunnel $NIS^{\prime }S$ junction takes place for
small transparencies, like in $NcS$ contacts \cite{BTK}. We note, that the
case of $SS^{\prime }cS^{\prime }S$ ballistic constrictions can be
considered in the same way, by extension of the OTBK model \cite{OTBK} with
the help of eqs.(15), (16) for $A(\epsilon )$ and $B(\epsilon )$ \cite
{Aminov}.

Therefore, the problem is reduced to calculation of the functions $%
\left\langle G_\epsilon (0+)\right\rangle $ and $\left\langle F_\epsilon
(0+)\right\rangle $ for the dirty $SS^{\prime }$ sandwich.

\section{Proximity effect in the dirty $SS^{\prime }$ sandwich}

The proximity effect in dirty $S^{\prime }S$ sandwiches was studied
previously in \cite{KL,GK-89} for the case of arbitrary transparency of the $%
S^{\prime }/S$ interface, where the case of a thin $S^{\prime }$ layer was
considered. Now we first generalize the results of \cite{KL,GK-89} to
arbitrary $S^{\prime }$ layer thickness and then, using these solutions,
calculate the quasiparticle current of a $NcS^{\prime }S$ contact.

The angle averaged quasiclassical Green functions in the dirty $S^{\prime }S$
bilayer satisfy the equation:

\begin{equation}
i\epsilon F_{s^{\prime },s}(x)+\frac{D_{s^{\prime },s}}2\left( G_{s^{\prime
},s}(x)\frac{\partial ^2F_{s^{\prime },s}(x)}{\partial x^2}-F_{s^{\prime
},s}(x)\frac{\partial ^2G_{s^{\prime },s}(x)}{\partial x^2}\right) -\Delta
_{s^{\prime },s}(x)G_{s^{\prime },s}(x)=0,
\end{equation}
where we have omitted the angular brackets $\left\langle ...\right\rangle $
for the functions $G_{s^{\prime },s},F_{s^{\prime },s}$ in $S^{\prime }$ and
$S$ regions respectively. Here $D_{s^{\prime },s}$ is diffusion coefficient,
$\Delta _{s^{\prime },s}$ is the order parameter. To be more specific, we
shall discuss below the particular case of $\Delta _{s^{\prime }}=0$, i.e.
of $T_{c^{\prime }}=0$ ($NS$ sandwich). The generalization to the case of
nonzero $T_{c^{\prime }}$ is straightforward \cite{GK-93} and does not
change our results qualitatively. The latter case will be discussed
separately elsewhere.

It is convenient to rewrite eq.(18) using the notations $G(\epsilon ,x)=\cos
\theta (\epsilon ,x),F(\epsilon ,x)=\sin \theta (\epsilon ,x)$. Then eq.(18)
takes the following form in $N$ and $S$ regions:

\begin{equation}
\xi _{N,S}^2\theta _{N,S}^{^{\prime \prime }}(x)+i\epsilon \sin \theta
_{N,S}(x)+\Delta _{N,S}(x)\cos \theta _{N,S}(x)=0,
\end{equation}
with the boundary conditions at the $NS$ interface($x=0$)\cite{KL}:
\begin{equation}
\begin{array}{c}
\gamma _B\xi _N\theta _N^{\prime }=\sin (\theta _S-\theta _N), \\
\gamma \xi _N\theta _N^{\prime }=\xi _s\theta _S^{\prime },
\end{array}
\end{equation}
in the bulk of the $S-$layer

\begin{equation}
\theta _s(\infty )=\arctan (i\Delta _0(T)/\epsilon ),
\end{equation}
as well as at the $N$ metal free surface ($x=-d_N$):

\begin{equation}
\theta _N^{^{\prime }}(-d_N)=0.
\end{equation}
The selfconsistency equation for the order parameter in $S$ region has the
form:

\begin{equation}
\Delta _s(x)\ln \frac T{T_c}+2\frac T{T_c}\sum_{w_n}\left[ \frac{\Delta
_s(x) }{\omega _n}-\sin \theta _s(x,\epsilon =i\omega _n)\right] =0.
\end{equation}
The parameters $\gamma _B$ and $\gamma $

\begin{equation}
\gamma _B=\frac{R_B}{\rho _N\xi _N},\gamma =\frac{\rho _S\xi _S}{\rho _N\xi
_N}
\end{equation}
have simple physical meanings: $\gamma $ is a measure of the strength of the
proximity effect between the $S$ and $N$ metals, whereas $\gamma _B$
describes the effect of the boundary transparency between these layers. Here
$\rho _{N,S}$, $\xi _{N,S}=\sqrt{D_{N,S}/2\pi T_c}$ and $D_{N,S}$ are normal
state resistivities, coherence lengths and diffusion constants of $N$ and $S$
metals, respectively, while $R_B$ is the product of the resistance of the $%
NS $ boundary and its area. We have normalized $\epsilon $ and $\Delta (x)$
to $\pi T_c$ , where $T_{c\text{ }}$is the critical temperature of the bulk $%
S$.

Previously selfconsistent solutions of the boundary value problem (19)-(24)
for the $NS$ sandwich at arbitrary values of $\gamma $ and $\gamma _B$ were
studied only for a thin $N$ layer, $d_N\ll \xi _N$, in \cite
{GK-89,GK-93,GK-95}. In particular, the densities of states in $N$ and $S$
layers $N_{N,S}(\epsilon )=Re\left( \cos \theta _{N,S}(\epsilon )\right) $
were discussed. It was shown for this case that for any values of $\gamma $
and $\gamma _B$ a superconducting state is induced with finite energy gap $%
\Delta _{gN}.$ When both layers are thin $d_{N,S}\ll \xi _{N,S}$, the
results of the well known McMillan tunnel model \cite{mcm} of the proximity
effect can be reproduced \cite{GK-95}. Then, in the language of the McMillan
model, the existence of a finite gap in the considered diffusive limit is
due to finite inverse lifetime of quasiparticles in thin $N$ layer, $\Gamma
_N\equiv \tau _N^{-1}\sim (\hbar v_{FN}/2d_N)D$, where $D$ is the angle
averaged $NS$ boundary transparency. As a result, there is a finite minimal
energy of quasibound states in $N$ and therefore, a nonzero energy gap is
induced. For finite thickness $d_N$ the densities of states in a dirty $NS$
sandwich were discussed previously only in the framework of rigid boundary
conditions, i.e. for $\gamma /\gamma _B\ll 1,$ in \cite{Zai-90,Volk,VZK}.

In the general case of arbitrary $d_N,d_S,\gamma $ and $\gamma _B$ the
solutions should be determined selfconsistently together with the spatial
dependence of the order parameter from the selfconsistency equation. The
results depend essentially on the parameters $\gamma ,\gamma _B,$ but from
lifetime considerations it is clear qualitatively, that the above conclusion
about a finite gap should hold also for any finite thickness of $N$. Below
we will calculate the density of states in $N$ from the solution of
eqs.(19)-(24) and will demonstrate the reduction of the gap in $N$ with
increase of $d_N$.

Taking advantage of the condition $\Delta _N=0$ one can integrate the
eq.(19) in $N$ region and with the help of boundary conditions (20) obtain

\begin{equation}
\cos \theta _N(0)-\cos \theta _N(-d_N)=\frac{\sin {}^2(\theta _S(0)-\theta
_N(0))}{2i\epsilon \gamma _B}.
\end{equation}
The analytic solution of eq.(25) is simplified only in a limiting case of
small $\gamma /\gamma _B$ ratio, when one can take $\theta _S(0)=\theta
_S(\infty )$ in the first approximation. Then for sufficiently low energies $%
\epsilon \ll \gamma _B^{-1},$ it follows from (25) that $\theta _N(0)\simeq
\theta _S(0)=\arctan (i\Delta _0/\epsilon )$, and as a result we obtain the
BCS expression with a gap $\Delta _0$ for the density of states in this
energy range: $N_N(\epsilon ,0)=Re\left( \epsilon /\sqrt{\epsilon ^2-\Delta
_0^2}\right) .$ For large $N$ layer thickness, $d_N\gg \xi _N$, one can
substitute $\cos \theta _N(d)=1$ in eq.(25) and find the asymptotic behavior
of $N_N(\epsilon )$ at the $NS$ boundary:

\begin{equation}
N_N(\epsilon )=\sqrt{\frac \epsilon {\pi T_c}\gamma _B\text{ }},\text{ }%
\frac \epsilon {\pi T_c}\gamma _B\ll 1.
\end{equation}

For the case of arbitrary $N$ layer thickness and arbitrary values of the
parameters $\gamma ,\gamma _B$ the boundary value problem (19)-(24) was
solved numerically. The results of the calculations of the densities of
states in $N$ at $x=0$ and $x=-d_N$ at low temperatures $T\ll T_c$ are
presented in Figs.1 and 2 for a number of $d_N/\xi _N$ ratios. As is seen
from Fig.1 at $x=-d_N$ (free surface of $N$) two peaks exist in $N(\epsilon
) $ for small $\gamma $ values, provided that $d_N/\xi _N\leq 1$. The first
peak is at $\epsilon =\Delta _{gN}$ and the second one at $\epsilon =\Delta
_0$. It is also seen from comparison of Figs.1 and 2, that the second peak
at $\epsilon =\Delta _0$ is smeared out with the increase of $d_N$ as well
as with the increase of $\gamma .$ On the other hand, the peak at $\epsilon
=\Delta _{gN}$ becomes more pronounced at large $d_N$. Dotted lines show the
behavior of $N(\epsilon )$ in $N$ at the boundary with $S$ ($x=0$). The
asymptotic behavior given by eq.(26) should take place at large $d_N$. It is
important to note, that the energy gap $\Delta _{gN}$ is preserved for all $%
d_N$, going to zero rather slowly. This is consistent with the qualitative
picture of the gap being proportional to inverse lifetime in $N$, which in
the considered diffusive approximation is given by $\tau _N^{-1}\sim (\hbar
D_N/d_N^2)D$. The results of study of the dependence of $\Delta _{gN}$ on
the parameters of the $NS$ sandwich will be presented in more detail
elsewhere.

Using the solutions $\theta _N(\epsilon ,-d_N)$, one can calculate the
reflection coefficients $A(\epsilon )$ and $B(\epsilon )$ and then the
quasiparticle current for the $NcNS$ contact from eqs.(15)- (17). To be more
specific, let us discuss here the case of a thin $N$ layer, $d_N/\xi _N\ll 1$%
. Then the number of parameters is reduced from $\gamma ,\gamma _B$ and $d_N$
to the following set: $\gamma _m=\gamma d_N/\xi _N$ and $\gamma _{BN}=\gamma
_Bd_N/\xi _N$ \cite{KL}.

Fig.3 shows the results of the calculations of the reflection coefficients $%
A(\epsilon )$ and $B(\epsilon )$ for a number of $Z$ values. It is seen that
for $Z\neq 0$ a characteristic two-peak structure exists for both $%
A(\epsilon )$ and $B(\epsilon )$, which is directly related to the two-peak
structure of $N_N(\epsilon )$ discussed above, the first peak being at $%
\epsilon =\Delta _{gN}$ and the second one at $\epsilon =\Delta _0.$

The zero-temperature conductance of $NcNS$ junctions calculated according to
eq.(17) is shown in Fig.4 for the same parameters. Again the two-peak
structure is present in $dI(V)/dV$ at voltages $\Delta _{gN}$ and $\Delta
_0. $ In accordance with the arguments given above, the peak at $\Delta _0$
would be smeared out quite easily by any pair-breaking process, i.e. by
large $\gamma _m$ values (spatial gradients in $S$), or large $d_N$. Then
such a spectroscopy will show only a proximity induced energy gap in $N$
with almost no signatures of $\Delta _0.$ The position of the first peak at $%
eV=\Delta _{gN}$ can be used to study properties of the $NS$ interface for
any given material combination. The appearance of such a conductance peak at
low bias for $Z\neq 0$ is a consequence of the given model, which follows
directly from the structure of the densities of states in the $N$ region at $%
x=-d_N$, as is seen from Figs.1 and 2.

\section{ Conclusions}

In conclusion, the BTK model is generalized for a spatially inhomogeneous
case of $NcS^{\prime }S$ ballistic constrictions with disordered $S^{\prime
}S$ electrodes. The expressions for the amplitudes of Andreev and normal
reflection are given, which allow to calculate a quasiparticle current for
arbitrary parameters of $S^{\prime }$ and $S$ materials and their interface,
if the conditions of the dirty limit are fulfilled. An energy gap in $%
S^{\prime }$ is always present, even for finite thickness of the $S^{\prime
} $ layer. The magnitude of this gap is studied as a function of the
parameters of the $S$ and $S^{\prime }$ materials, as well as of the
transparency of the $SS^{\prime }$ interface. It is shown, that the
conductance of ballistic $NcS^{\prime }S$ junctions reflects proximity
induced energy gap in $S^{\prime }$ and, under certain conditions, also the
bulk gap of the superconductor $S.$

\vspace{2.0cm}

{\bf Acknowledgements}. Stimulating discussions with B.Aminov, D.Averin and
K.Likharev are gratefully acknowledged. Both authors are thankfull to
A.Braginski for hospitality during their stay at KFA-J\"ulich, where the
work was completed. The work was supported in part by the International
Science Foundation under Grant No.MDP000, Russian Ministry of Scientific and
Technical Policy in the frame of the Scientific Program ''Actual Problems of
Condensed Matter Physics'' and BMFT Germany (Grant 13N6329).

\begin{figure}
\caption{
The densities of states in the $N^{\prime }$ layer of the $N^{\prime }S$
sandwich, normalized to their normal-state
values, at the free surface (solid lines) and
at the $N^{\prime }S$ boundary
(dashed lines) for different thicknesses $d_{N}/\xi_{N}= 10$ (curve 1),
2 (2), 1 (3) and 0.5 (4).}
\label{Fig.1}\end{figure}

\begin{figure}
\caption{
The densities of states in the $N^{\prime }$ layer of the $N^{\prime }S$
sandwich, normalized to their normal-state
values, at the free surface (solid lines) and
at the $N^{\prime }S$ boundary
(dashed lines). Thicknesses $d_{N}/\xi_{N}$ are the same as in Fig.1.}
\label{Fig.2}\end{figure}

\begin{figure}
\caption{
Probabilities of Andreev reflection, $A(\epsilon)$ (solid lines),
and of normal reflection, $B(\epsilon)$ (dashed lines),
for the ballistic  $NcN^{\prime }S$ constriction with $\gamma_m = 0.1$
and $\gamma_{BN} = 1$.}
\label{Fig.3}\end{figure}

\begin{figure}
\caption{
Zero-temperature conductance of the ballistic $NcN^{\prime }S$ constriction
with the same parameters as in Fig.3.}
\label{Fig.4}\end{figure}

\end{document}